\documentclass[%
 aip,
 amsmath,amssymb,
superscriptaddress]{revtex4-1}

\usepackage{graphicx}
\usepackage{dcolumn}
\usepackage{bm}

\usepackage[utf8]{inputenc}
\usepackage[T1]{fontenc}
\usepackage[dvipsnames]{xcolor}
\usepackage{mathptmx}
\usepackage{amsmath}
\usepackage{etoolbox}
\usepackage{physics}
\usepackage{charter}
\usepackage{tempora}
\DeclareUnicodeCharacter{2014}{\sar}
\makeatletter
\def\@email#1#2{%
 \endgroup
 \patchcmd{\titleblock@produce}
  {\frontmatter@RRAPformat}
  {\frontmatter@RRAPformat{\produce@RRAP{*#1\href{mailto:#2}{#2}}}\frontmatter@RRAPformat}
  {}{}
}%
\makeatother
\begin{document}


\title{Elasticity can affect droplet coalescence}
\author{Sarath Chandra Varma}
 \affiliation{%
$^1$Department of Mechanical Engineering, Indian Institute of Science, Bangalore, Karnataka-560012, India
}%

\author{Debayan Dasgupta}%
\affiliation{%
$^1$Department of Mechanical Engineering, Indian Institute of Science, Bangalore, Karnataka-560012, India
}%
\author{Aloke Kumar*}
 \email{alokekumar@iisc.ac.in}
 \affiliation{%
$^1$Department of Mechanical Engineering, Indian Institute of Science, Bangalore, Karnataka-560012, India
}%


\begin{abstract}
Recent investigations on the coalescence of polymeric droplets on a solid substrate have reported strong disagreements; the heart of the issue is whether coalescence of polymeric drops is similar to that of Newtonian fluid and is independent of molecular relaxation, or whether the role of entanglement of polymeric chains leads to a transition kinetics different from that of Newtonian fluid. Via this report, we resolve the disagreements through a discussion on the effects of merging method on the dominant forces governing the coalescence process, i.e., inertia, dissipation, and relaxation. Our study unveils that the coalescence dynamics of polymeric drops is not universal and in fact, it is contingent of the method by which the coalescence is triggered. Additionally, we demonstrate the spatial features of the bridge at different time instants by a similarity analysis. We also theoretically obtain a universal bridge profile by employing the similarity parameter in a modified thin film lubrication equation for polymeric fluids.
\end{abstract}

\maketitle

\section{Introduction}
Coalescence of droplets on a solid surface, also known as sessile sessile coalescence, is key to a number of commercial applications including mixing of reagents in microfluidics system \cite{microfluidic}, inkjet printing \cite{inkjetprinting}, electronic packaging \cite{electronicpackaging} and rapid prototyping \cite{rapidprototyping}. It involves an initial rapid growth of meniscus bridge, followed by a slow rearrangement of the droplets shape from elliptical to spherical cap at longer times. The presence of solid substrate in such configurations slows down the liquid transport towards the bridge and imposes additional challenges of capturing the complex contact line motion and energy interaction between solid, liquid and vapour phases. Once the initial contact is developed between the droplets, the droplet contour is described by the evolution of bridge height, $h_b$, perpendicular to the substrate and bridge width, $r_m$, parallel to the substrate. In case of Newtonian fluids,  potential technological interest has driven a lot of effort in investigating the effects of surface wettability \cite{surfacewettability1,  surfacewettability2,  surfacewettability3}, viscosity \cite{viscosity}, droplet size \cite{dropletsize,  dropletsize2} and contact angle hysteresis \cite{contactanglehystereris, contactanglehystereris2} on the growth of meniscus bridge at both initial and later stages of the coalescence. In particular, Narhe et al. \cite{narhe2008dynamic} proposed a scale of $h_b\sim t$ and $r_m\sim t^{1/2}$ for initial stages of coalescence, where capillary number (Ca) was greater than 0.2. Ristenpart et al. \cite{PRL2006} also reported an exponent of 1/2 for growth of the meniscus bridge width on a highly wettable surface. A deviation from the proposed scale of $t^{1/2}$  was observed by Leet et al. at higher contact angles and longer times \cite{lee2012coalescence}. However, at initial stages of coalescence, they identified a power law exponent for the bridge height that ranged between 0.5 to 0.86 and increased with increase in contact angle. Hernandez et al. \cite{PRL2012} revealed that the bridge height grew linearly with time and evolved with a self similar dynamics. In addition to the slow viscous regime discussed so far, the bridge height was observed to grow with a universal exponent of 2/3 for contact angle below $90^\mathrm{o}$ and an exponent of 1/2 for contact angle of $90^\mathrm{o}$ in the inertial  regime \cite{sui2013inertial, PRL2013, pawar2019symmetric}.

In contrast to Newtonian droplets discussed above, the coalescence of rheologically complex fluids rather remains obscure despite its wide application in droplet 3D printing \cite{van2014uv, klestova2019inkjet}, emulsions \cite{emul1, emul3} and microfluidics \cite{krebs2012coalescence}. Varma et al. \cite{our} highlighted the importance of viscoelasticty and relaxation time for coalescence of polymeric drops in a pendant-sessile configuration. A scale of $r\sim t^{0.36}$ was reported for neck growth, which is a significant deviation from $r\sim t$ and $r\sim t^{1/2}$ observed for Newtonian droplets\cite{paulsen2011viscous,xia2019universality} in the viscous and inertial regime, respectively. However, at very high concentrations, Varma et al. \cite{varma2022rheocoalescence} showed a continuous decrease in power law index from 0.36. This is further supported qualitatively by a numerical study on polymers and microgels  by Chen et al.\cite{chen2022viscoelastic}. Even a seperate study for coalescence of polymeric droplet in sessile sessile configuration by Varma et al.\cite{varma2021coalescence} reported  a decrease in exponent from 2/3 in the inertial regime to 1/2 in the viscoelastic regime. Correspondingly,  a recent numerical study by Chen et al.\cite{POF2022} investigating coalescence of non-elastic, shear-thinning fluid highlighted a strong relation between power law rheology and scaling exponent at the onset of coalescence. In this regard, it is worth keeping in mind that macromolecular fluids often exhibit strong shear thinning characteristics as well. Interestingly, experimental assessment by Dekkar et al. \cite{dekker2022elasticity} highlighted that the presence of polymers causes negligible effect on the temporal evolution of bridge height. This was supported by the observation that a wide range of polymeric concentrations reported a universal power law index of 2/3, which is similar to that of DI water.  However, the kinematics of coalescence and pinching are not disparate; in fact, Fardin et al. \cite{fardin2022spreading} revealed the shared and universal features of these flows by showing an excellent collapse of experimental data pertaining to pinching, spreading and coalescence of Newtonian fluids into a universal scale. Interestingly, it is well promulgated that even a minute addition of polymer drastically alters the breakup dynamics of droplets \cite{clasen2006beads, eggers2020self} by inhibiting pinch off. In this regard, the conclusion by Dekker et al. \cite{dekker2022elasticity} that droplet coalescence is independent of complex fluid rheology seems counterintuitive and demands further exploration. A closer look reveals that the experimental methods adopted by Varma et al.\cite{varma2021coalescence} and Dekker et al.\cite{dekker2022elasticity} are different. Varma et al. \cite{varma2021coalescence} developed the initial contact by creating two pendant droplets of constant volume very close to each other, such that once the drops touched the substrate, they spread to achieve thermodynamic equilibrium. Spreading droplets can create a liquid bridge, as shown in Fig. 1a. Such coalescence mimics the scenario encountered in ink-jet printed liquid lines and electronic packaging, which require accurate placement of pendant polymeric droplets so that they spread and merge after impacting the substrate \cite{duineveld2003stability}. In the present text, this method is referred to as Droplet Spreading Method (DSM). In a different method as adopted by Dekkar et al. \cite{dekker2022elasticity}, two adjacent droplets are grown simultaneously by increasing the volume until the edges contact and coalescence take place. Such phenomenon is analogous to coalescence due to condensation \cite{somwanshi2018dropwise}. This method of merging is referred to as Volume Filling Method (VFM) and is represented in Fig. 1b. The continuous influx of fluid by VFM leads to a coalescence dynamics different from that of DSM. The influence of the two methods of spreading on bridge evolution of Newtonian drops has been demonstrated by Sellier et al. \cite{sellier2009modeling} through numerical modelling and experimental investigation. The neck growth predicted by experiment was observed to be two to three times larger than that of numerical model. The difference was attributed to the type of merging mechanism; while the merging was due to surface tension induced by capillarity in the numerical simulation, it was induced by volume growth in the experiment.  

Available information in literature shows that the effects of merging process on sessile sessile coalescence of polymeric drops is yet to be addressed. In particular, VFM coalescence of polymeric droplets warrants more discussion regarding the behavior of polymeric chains under continuous pumping.  Here, we present an experimental and theoretical investigation of coalescence of two symmetrical polymeric droplets on a substrate by VFM and compare the results with DSM. By introducing appropriate scaling parameters, we obtain a representative universal shape of the bridge near the meniscus in the powerlaw regime. Polyethylene oxide (PEO) of varying concentrations is chosen as the representative polymeric fluid and the coalescence is carried out on an aluminum substrate. Further, we attempt to theoretically demonstrate the universal shape of the bridge by employing the similarity parameter in a modified thin film lubrication equation in the semi-dilute entanglement regime. We believe that the present study will compliment the work on coalescence by spontaneous spreading \cite{varma2021coalescence} and make the overall research on sessile-sessile coalescence of polymeric drops more complete and comprehensive.

\begin{figure*}[htb]
\includegraphics[width=1\textwidth]{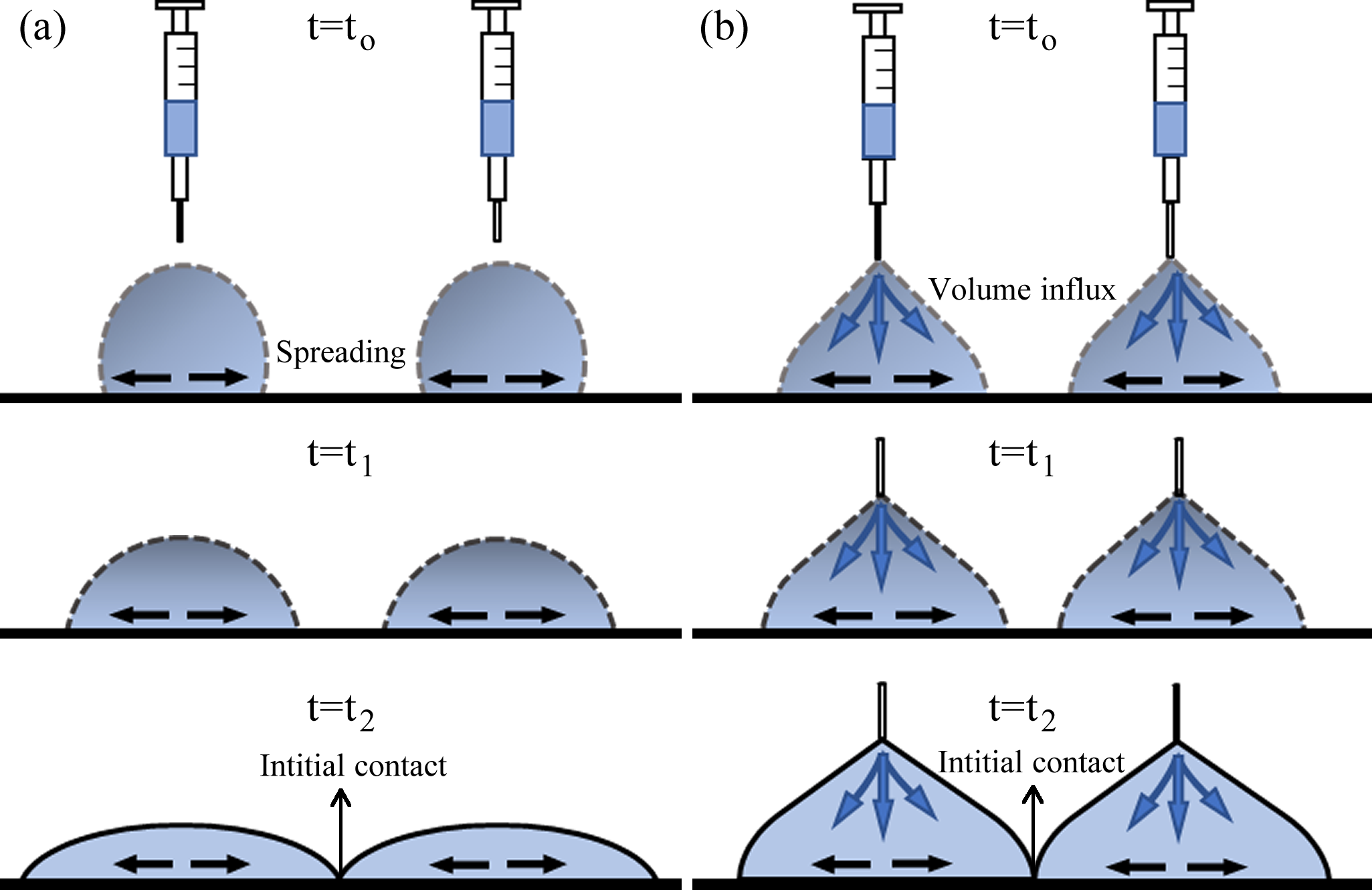} 
\caption{Merging of droplets on substrate by (a) Depositing fixed volume droplets and allowing them to spread to achieve coalescence (DSM), (b) Continuous pumping of droplets till the instance of coalescence (VFM). }
\label{fig:Schematic} 
\end{figure*}

\section{Materials and methods}
\subsection{Materials}
Consistent with Varma et al. \cite{varma2021coalescence}, we prepare ten solutions of concentrations $c$ (w/v) 0.01\%, 0.02\%, 0.05\%, 0.061\%, 0.1\%, 0.2\%, 0.3\%, 0.4\%, 0.5\% and 0.6\% by dissolving PEO with molecular weight $M_w= 5\times10^6$ g/mol (Sigma-Aldrich) in de-ionized (DI) water. In addition, we prepare another four solutions of sufficiently high concentrations 1.0\%, 1.5\% 1.75\% and 2.0\% having the same molecular weight. The homogeneity of the solutions is ensured by stirring them at 300 rotations per minute for at least 24 hours. The chosen concentrations belong to the dilute ($c/c^*<1$), semi-dilute unentangled ($1<c/c^*<c_e/c^*$), and semi-dilute entangled  ($c/c^*>c_e/c^*$) regime, where, $\displaystyle c^*$ is critical concentration and   $c_e$ is entanglement concentration.
 
Measurement of surface tension, $\sigma$, by pendant drop method using optical contact angle measuring and contour analysis systems (OCA25) instruments from Dataphysics yields surface tension values of $0.063 \pm 0.02 \mathrm{~N} / \mathrm{m}$ for all concentration ratios $(c/c^*)$. Density of the solutions obtained by measurement of mass and volume falls in the range of $1000 \pm 50 \mathrm{~kg} / \mathrm{m}^{3}$. Hence, a constant density of $1000 \mathrm{~kg} / \mathrm{m}^{3}$ has been assumed for all the polymeric solutions.

To achieve coalescence by VFM, a substrate with higher contact angle is desirable, as it resists spontaneous spreading. Hence, aluminium substrate (RS Components \& Controls (India) Ltd.) of dimension $80\times30\times1.25$ mm is used in the present case, as it displays a high contact angle. The substrates are first cleaned with detergent and then sonicated with acetone and water for 20 minutes each. Subsequently, they are placed in the oven at 95$^\circ$C for 30 minutes. Droplet geometry measured using ImageJ DropSnake toolbox shows that an interfacial contact angle of $72^\mathrm{o}\pm3^\mathrm{o}$ and overall contact length of $2R_0\approx 3 \pm 0.25$ mm is maintained for all concentrations of PEO solution.

\begin{table}[hbt!]
\caption{Rheological properties of the solutions.}
\label{tab:Table. S1}
\begin{ruledtabular}
\begin{tabular}{ccccc}
$c$ (\%w/v)&Concentration ratio($c/c^*$)&$\eta_o$ (mPa.s)&$\lambda$ (ms)\\
\hline
0 & - & 1 & DI Water\\
0.01 & 0.16 & 1.3 & 1.5\\
0.02 & 0.32 & 1.5 & 1.5\\
0.05 & 0.82 & 2 & 1.5\\
0.061 & 1 & 3 & 1.5\\
0.1 & 1.6 & 6 & 2\\
0.2 & 3.9 & 18 & 2.7\\
0.3 & 5 & 46 & 3.5\\
0.4 & 6.5 & 60 & 74\\
0.5 & 8.2 & 200 & 115\\
0.6 & 9.8 & 500 & 160\\
1.0 & 16 & 4.5 & 500\\
1.5 & 25 & 20 & 670\\
1.75 & 29 & 40 & 1325\\
2.0 & 33 & 55 & 1350\\

\end{tabular}
\end{ruledtabular}
\end{table}
\subsection{Experimental Setup}
\begin{figure*}[!ht]
\includegraphics[scale=0.75]{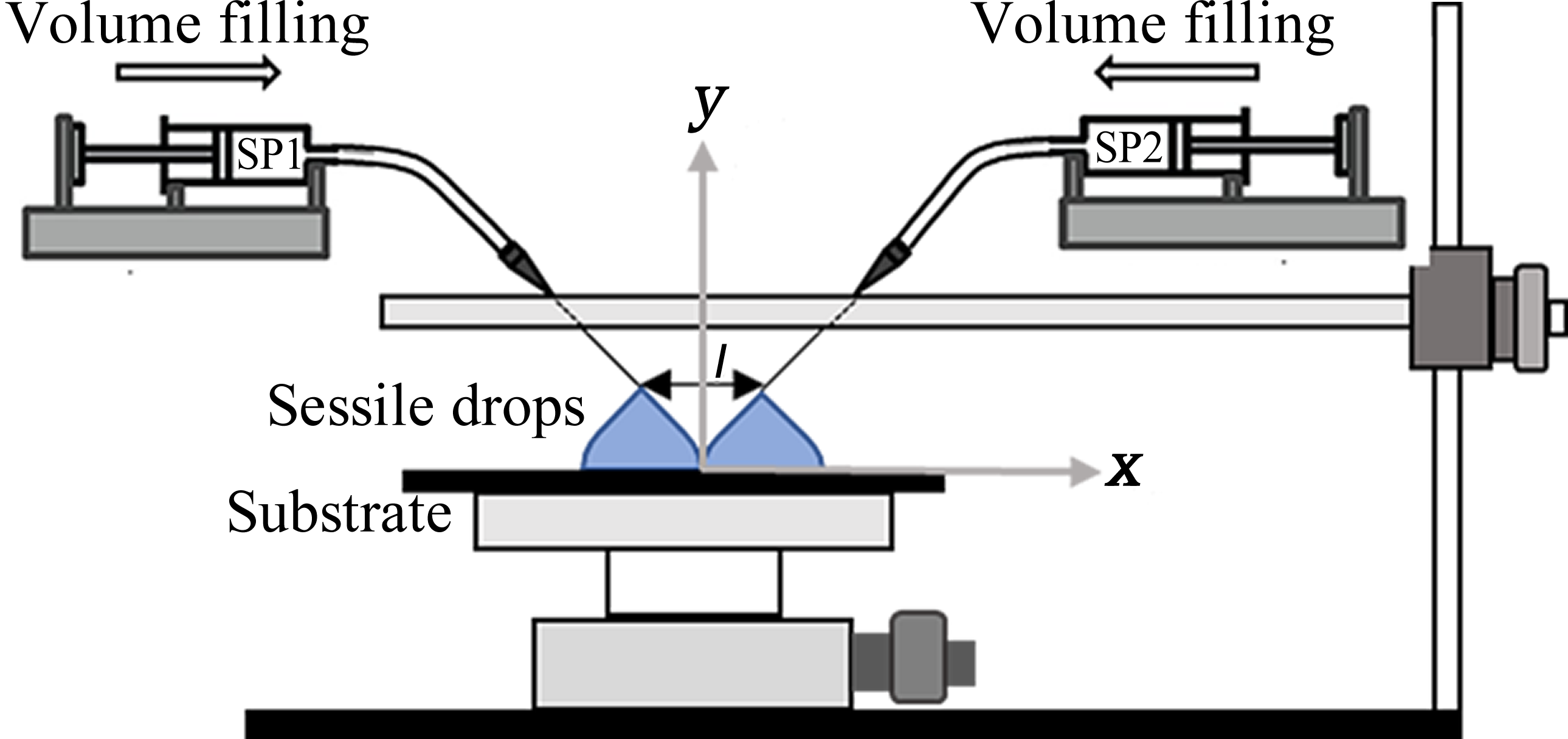} 
\caption{Schematic of the front view of the experimental setup (SP1 and SP2 are syringe pumps) for VFM}
\label{fig:setup} 
\end{figure*}

A fixture having a hollow cylinder with  a $45^\mathrm{o}$ tilted axis is 3D printed for holding the needles. The needles are inserted through top of the fixture as shown in Fig.~\ref{fig:setup}. Two symmetric drops are grown at the tip of two flat Nordson needles with  0.41 mm inner and 0.71 mm outer diameter. The tip of the needles coming out from the other end are separated by a distance of $l\approx 2$ mm. Initial distance between needle tip and substrate is kept at 0.55 mm. Once the droplets reach a volume $\sim $ 3 $\mu\mathrm{l}$, they touch the substrate. The substrate is further lowered very slowly by a distance of 0.85 mm to ensure that the meniscus connecting the needles to the droplet does not affect the merging process. Finally, coalescence is achieved by quasi-statically advancing the contact lines towards each other due to continuous injection pumping. A small pumping rate of 2 $\mu\mathrm{l/min}$ throughout the process ensures that the droplets are in their equilibrium shape at all time. The liquid supply is stopped as soon as the drops come in contact with each other and the dispensed volume is noted. Moreover, additional experiments are also conducted for  $c/c^*$=16, 25 29 and 33 for DSM, following the experimental method of Varma et al. \cite{varma2021coalescence}. A 45W LED light source (Nila Zaila, USA) at 100\% output is used for backlight diffusive illumination of the region of coalescence. A Photron Fast-cam mini AX-100 high speed camera coupled with Navitar 6.5x zoom lens records the whole process at 60,000 frames per second and 1/100,000 s shutter speed. 
The shape and evolution of the interface is tracked using a sub-pixel based edge detection algorithm in MATLAB. At first, the images are binarized using an appropriate thresh hold pixel intensity to get rid of the background disturbances.  The coordinate of the column at the contact point of two drops represent the position of the bridge. By tracking the pixel intensities along the column, the row where the pixel intensity falls below the set threshold pixel value is noted. The coordinate values thus obtained are utilized to get the surrounding grey intensities in original frame. A pixel weighted average method is implemented to obtain the sub pixel coordinates of the bridge. These coordinates are finally subtracted from the substrate coordinates to obtain the bridge height. The process is repeated for each frame to extract the evolution of the bridge height with time. Similarly, the evolution of the bridge profile $h(x,t)$ in the proximity of $h_b$ is obtained by tracking 15 columns on either side of $h_b$ using the same procedure for each frame.    

\section{Rheology: Critical concentration and relaxation time}
The critical concentration, $\displaystyle c^*$, and entanglement concentration, $\displaystyle c_e$, of the PEO solutions are represented by $\displaystyle c^*=1/[\eta]$ and $\displaystyle c_e\approx 6c^*$ \cite{arnolds2010capillary}, respectively, where, the intrinsic viscosity $\eta$ is obtained from Mark-Houwink-Sakurada correlation\cite{tirtaatmadja2006drop} $[\eta]=0.072M_w^{0.65}$. For molecular weight $M_w= 5\times10^6$ g/mol, a critical concentration value of $\displaystyle c^*=0.061\%$ w/v and entanglement concentration of $c_e=0.366\%$ w/v is thus obtained. 
The relaxation time, $\lambda$, in the dilute regime is estimated using the Zimm model \cite{bird1987dynamics}.

\begin{equation}
  \lambda_z=\frac{1}{\zeta(3\nu)}\frac{[\eta]M_w\eta_s}{\mathrm{N_A k_B} T}   
\end{equation}

where, $\displaystyle \lambda_z$ is the Zimm relaxation time, $\eta_s$ is the solvent viscosity, $\mathrm{k_B}$ is the Boltzmann constant, $\mathrm{N_A}$ is the Avogadro number, $T$ is the absolute temperature, $a$ is the exponent of Mark-Houwink-Sakurada correlation and $\nu$ is fractal polymer dimension obtained from $a=3\nu -1$. The relaxation time in semi-dilute unentangled and and semi-dilute entangled regimes are represented by $\displaystyle \lambda_{\mathrm{SUE}}$ and $\displaystyle \lambda_{\mathrm{SE}}$, respectively, and are calculated using the correlations $\displaystyle \lambda_{\mathrm{SUE}}=\lambda_z\Big(\frac{c}{c^*}\Big)^{\frac{2-3\nu}{3\nu-1}}$ and $ \displaystyle \lambda_{\mathrm{SE}}=\lambda_z\Big(\frac{c}{c^*}\Big)^{\frac{3-3\nu}{3\nu-1}}$ 
\cite{rubinstein2003polymer,liu2009concentration,del2015rheometry}. The relaxation times and concentration ratios, $c/c^*$, corresponding to the chosen concentrations in the study are listed in Table-1. Viscosity, $\eta_o$, of the solutions given in Table-1 are obtained from Varma et al.\cite{varma2021coalescence}

\begin{figure}[!h]
\includegraphics[width=0.6\textwidth]{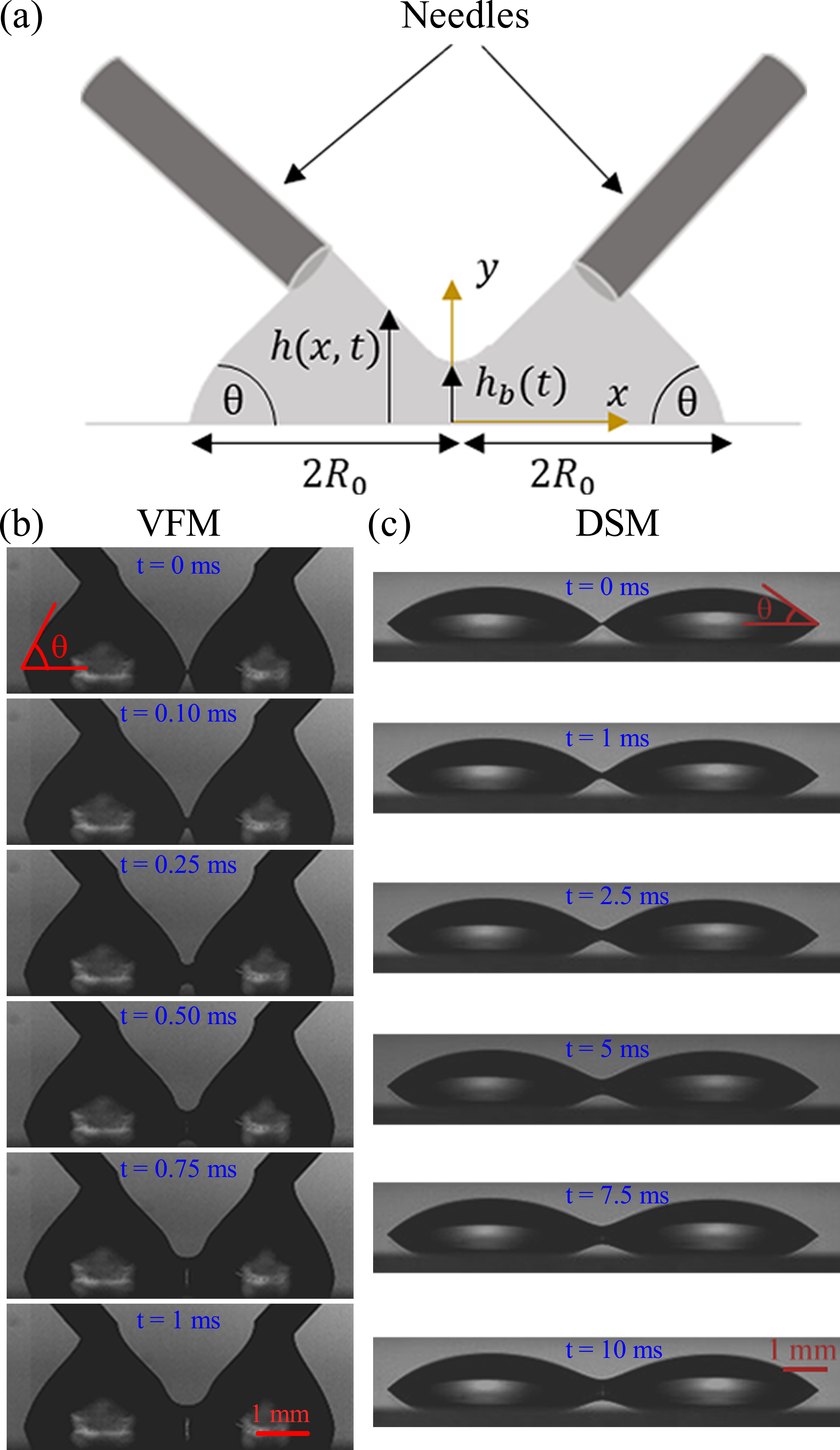} 
\caption{(a) Schematic of the coalescence representing the geometric parameters for VFM. Snapshots showing the bridge evolution of  0.5 \% w/v concentration ($c/c^*$=8.2) of PEO at different instants of time by (b) VFM (c) DSM}
\label{fig:cascade} 
\end{figure}

\section{Results and Discussion}
The initial contact between droplets is obtained by quasi-statically increasing the droplet volumes at a pumping rate of 2 $\mu$l/min. This ensures that the approach velocity is negligible and the droplets are in thermodynamic equilibrium at all times. At the beginning of the coalescence, a tiny liquid bridge develops at the point of contact between the droplets. The large radius of curvature of the bridge results in fluid flow from the close neighbourhood towards the bridge region due to capillary action. This fluid flux may disturb the equilibrium at the pinned ends at intermediate stage of coalescence when the bridge relaxes, causing a change in contact angle. However, the time taken for the disturbance to reach the pinned end is much higher than the time scale of interest for the present case. Hence, the contact angle is considered to be constant for all practical purposes.  

Fig.~\ref{fig:cascade}(a) shows the schematic of the drop coalescence, where $2R_0\approx 3 \pm$0.25 mm represents the contact length and  $\theta\approx72\pm3^{\circ}$ represents the interfacial contact angle. Evolution of the liquid bridge during the coalescence of two PEO droplets with concentration ratio $c/c^*=8.2$ obtained by VFM and DSM are shown in Fig.~\ref{fig:cascade}(b) and (c), respectively, at different instants of time.

\begin{figure}[h]
\includegraphics[width=1\textwidth]{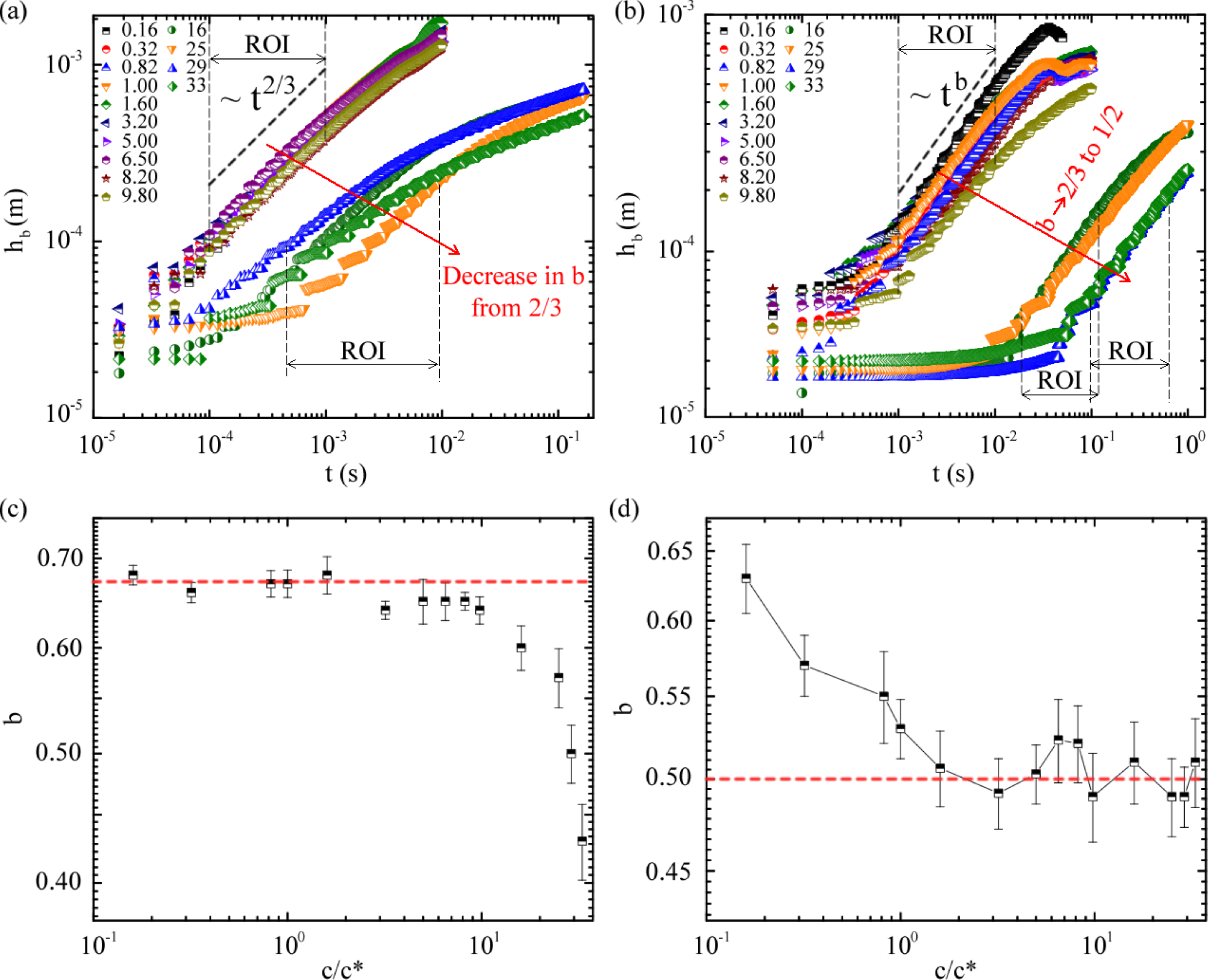} 
\caption{(a) Temporal evolution of bridge height by VFM showing a constant power law index $b$ of 2/3 for $c/c^*\le 9.8$ and monotonous reduction with $c/c^*$ for $c/c^*>9.8$  (b) Temporal evolution of bridge height by DSM obtained from Varma et al.\cite{varma2021coalescence} along with experimentally obtained data for four additional concentrations ($c/c^*$=16, 25 29 and 33). showing a decrease in exponent $b$ from 2/3 to 1/2 with increase in $c/c^*$ (c) Variation of exponent $b$ with $c/c^*$ for  VFM (d) Variation of exponent $b$ with $c/c^*$ for DSM}
\label{fig:hvst} 
\end{figure}

Fig.~\ref{fig:hvst}(a) and (b) shows the temporal evolution of the bridge height $h_b=h(0,t)$ obtained by VFM and DSM, respectively, for the complete range of $c/c^*$. The bridge height values are the average of 5 trials conducted for each solution. The experiments are extremely repeatable and the error in measurement is limited to $\pm5\%$. It can be observed that the regions of interest (ROI) corresponding to the early power law regime is at relatively smaller times scales for VFM as compared to DSM for the same polymer concentration ratios. In case of VFM, Fig.~\ref{fig:hvst}(a) shows that the exponent $b$ of the power law growth registers a constant value of 2/3 for $c/c^*\le9.8$, beyond which it continuously reduces with increase in $c/c^*$. For the considered range of $c/c^*$, $b$ does not seem to achieve any stable value. On the other hand, when coalescence is triggered by droplet spreading (Fig.~\ref{fig:hvst}(b)), the presence of polymer clearly affects the growth of the meniscus bridge even at small polymer concentrations ($c/c^*<1$). This is highlighted by a reduction in the exponent of growth from 2/3 for $c/c^*<1$ to a constant value of 1/2 for $c/c^*>1$. Correspondingly, the exponent $b$ deduced by fitting the power law curve $h_b\sim t^{b}$ is shown in Fig.~\ref{fig:hvst}(c) and (d) for VFM and DSM, respectively. The apparent contrast in the flow kinetics for the two methods suggest that the effect of polymers on coalescence is strongly influenced by the experimental method. A qualitative discussion on the behaviour of polymer chains subjected to the two methods throws some light on the associated disparity. When the droplets are maneuvered towards each other by spontaneous spreading through DSM, the polymeric chains get elongated and attain an unrelaxed state. As polymeric concentration increases, these unrelaxed polymer chains offer stronger resistance to the growth of the bridge in the vertical direction, which results in a decrease in the growth exponent, as observed in the case of Varma et al. \cite{varma2021coalescence}. However, in the present case of VFM, where the droplets approach each other due to continuous influx of liquid by pumping, the polymeric chains do not undergo enough elongation and mostly remain relaxed. Subsequently, the resistance offered by the polymeric chains to the initial bridge growth is relatively weak. Moreover, the effect of inertial forces induced by continuous pumping in VFM also needs to be considered.  Notably, Dekker et al. \cite{dekker2022elasticity} also observed a continuous reduction in exponent $b$ with $c/c^*$ at higher values of $c/c^*$ but contributed the same to error arising from determining the initial point of coalescence. However, in addition to the qualitative explanation provided above, we address this phenomenon by identifying the  dominance of the underlying forces through two time averaged non-dimensional numbers - Reynolds number, $Re=<\rho u_cl_c/\eta_o>$, and the Weissenberg number, $Wi=<\lambda u_c/l_c>$, where $\eta_o$ is the zero shear viscosity, $u_c\sim\partial h_b/\partial t$ is the characteristic velocity scale and $l_c\sim h_b$  is the characteristic length scale.
\begin{figure}[ht]
\includegraphics[width=1\textwidth]{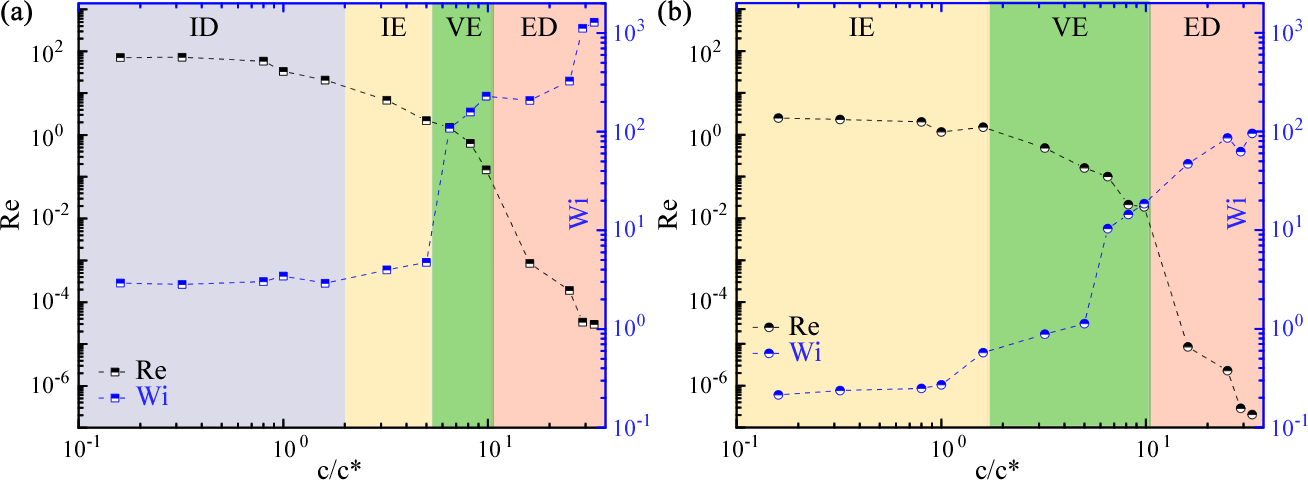} 
\caption{Variation of Reynolds number $Re$ and Weissenberg number $Wi$ with $c/c^*$ for (a) VFM and (b) DSM}
\label{fig:numbers} 
\end{figure}

Fig.~\ref{fig:numbers}(a) and (b) shows a comparison between $Re$ and $Wi$ values obtained by the present VFM method and that obtained via DSM by Varma et al. \cite{varma2021coalescence}, respectively. 
In the subsequent discussion, the inertial, viscous and elastic forces are represented by $F_i$, $F_v$ and $F_e$, respectively. At first, we consider the influence of $c/c^*$ on $Re$ and $Wi$ for VFM. It can be observed that $Re\sim {O}(10^2)$ and $Wi\sim {O}(10^0)$ for $c/c^*<1.6$. Hence, the sequence of the participating forces is described as $F_i>>F_v\sim F_e$ and the corresponding regime is identified as an inertia dominated regime (ID). For $1.6<c/c^*<c_e/c^*$, $Re$ ranges between ${O}(10^1)$ and ${O}(10^0)$, and $Wi$ approaches ${O}(10^1)$, thereby representing an inertio-elastic regime (IE), where, $F_e>F_i>F_v$. In the subsequent region $c_e/c^*<c/c^*<10$, a viscoelastic regime (VE) is identified, where $Re \sim {O}(10^{-1})$ and $Wi \sim {O}(10^2)$, such that $F_e>F_v>F_i$. At sufficiently high polymer concentrations $c/c^*>10$, $Re$ approaches ${O}(10^{-5}$) and $Wi$ approaches ${O}(10^3)$, which suggests that both the inertial and viscous forces become negligible and the flow dynamics is completely taken over by the elastic forces. This results in an elasticity dominated regime (ED) characterized by $F_e>>F_v>>F_i$. A similar comparison between $Re$ and $Wi$ values for DSM in Fig.~\ref{fig:numbers}(b) reveals three regimes namely inertio-elastic, viscoelastic and elasticity dominated regimes. A regime-wise comparison between the two methods reveals that the inertial forces are more prominent in case of VFM, as compared to DSM. In fact, no inertia dominated regime is observed for DSM. At most the inertial forces are comparable with the elastic forces for $c/c^*<1.6$ and become trivial beyond that. As a result, the flow dynamics is inertia dominated in VFM, whereas it is inertio-elastic in DSM for $c/c^*<1.6$. Consequently, the elastic forces in VFM cannot surpass the strong inertial forces and the effect of polymer remains quiescent. On the other hand, the weak presence of inertial forces in this regime for DSM coupled with spreading induced polymeric chain elongation causes a reduction in the exponent of growth of the bridge height. Interestingly, Sellier et al. \cite{sellier2009modeling} also observed a faster neck growth for the volume growth method as compared to spreading for Newtonian fluid, suggesting a relatively stronger effect of inertial forces for volume growth induced merging. Thus, VFM is inherently associated with stronger inertial forces, which results in higher exponent of growth in the absence or presence of weak elastic forces. Only at sufficiently large polymer concentrations ($c/c^*>9.8$), the polymer chains get highly entangled and a notable effect of elasticity is observed. It is also evident from the above discussion that the thin film equation can only be applied for $c/c^*>9.8$, where the effect of inertial forces are less.   

We suggest that similar to Newtonian fluids, early stages of coalescence of polymeric drops can also be characterized by a self similar meniscus profile. However, a scaling parameter same as that of Newtonian fluids can not be chosen due to the presence of an inherent relaxation time scale in the polymeric fluids. In fact, Varma et al. \cite{varma2021coalescence} showed that the relaxation time scale $(\lambda)$ is the most important nodal parameter that governs the power law behavior of the polymeric liquid. Hence, the self similar  parameter is perturbed with the Weissenberg number, $Wi$, to accommodate the inherent time scale of polymers into the similarity parameter. The new self similar regime for polymeric coalescence is expressed as
\begin{equation}
 h(x,0)=h_0(t) \zeta (\xi),  \xi=\frac{\theta x}{2vt}\left(1+\frac{1}{1+Wi}\right)  
\label{eq:similarity}
\end{equation}
where, $\zeta (\xi)$ is the similarity profile of the bridge. Substitution of $Wi=0$ in Eq.~\eqref{eq:similarity} reduces the similarity variable to $\xi=\frac{\theta x}{v t} $, which is the similarity parameter adopted by Hernandez et al.\cite{PRL2012} for Newtonian fluid.

\begin{figure*}[h!]
\includegraphics[scale=0.38]{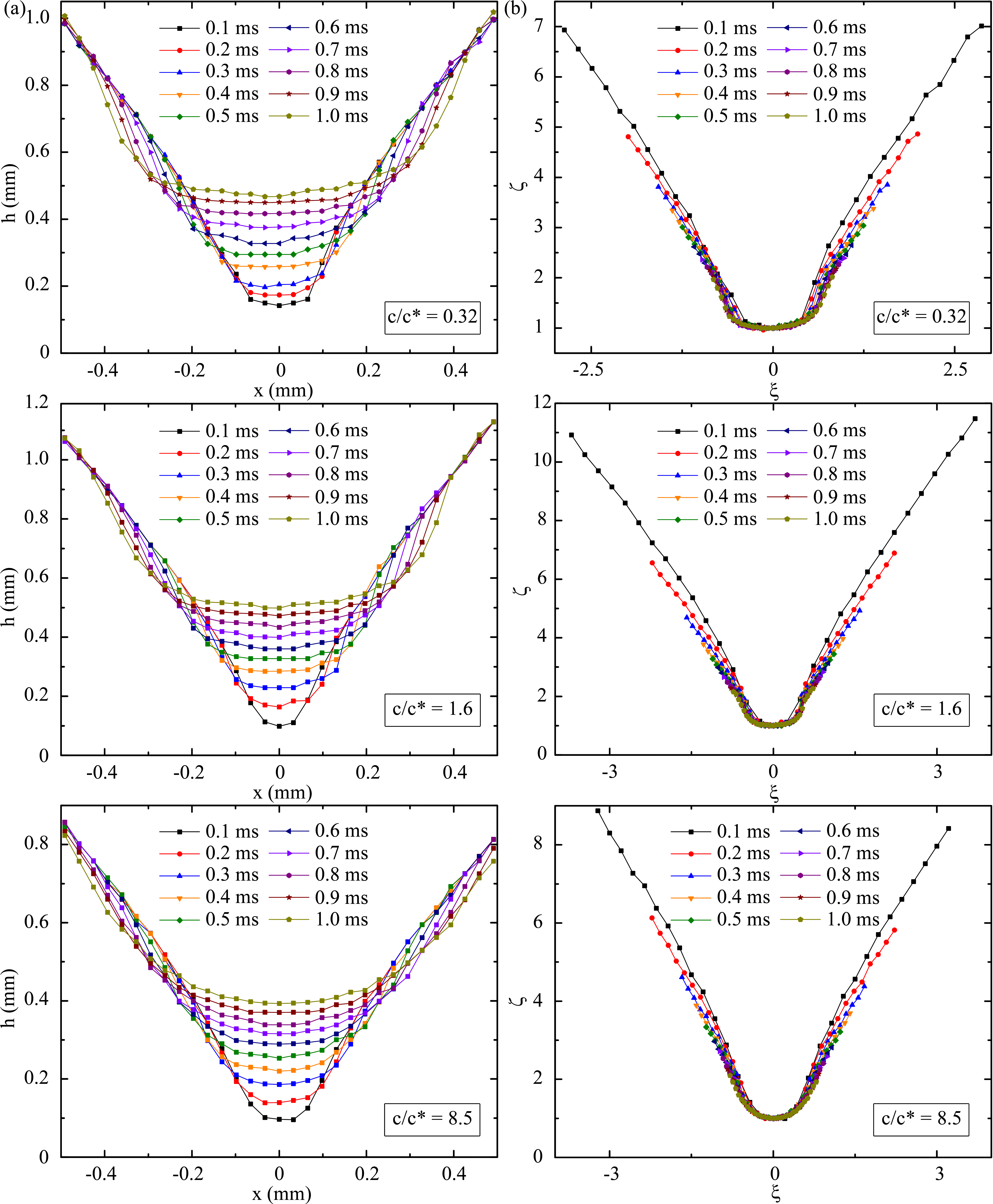} 
\caption{(a) Bridge profile for various concentration ratios $c/c^*$ at different time instants ranging between 0.1 ms to 1.0 ms with an interval of 0.1 ms  (b) Re-scaled bridge profiles revealing a self similar dynamics through collapse of data in the same time range}
\label{fig:computational} 
\end{figure*}

Fig.~\ref{fig:computational}(a) shows the bridge profile at different instants of time for three polymeric concentrations $c/c^*=0.32$, $c/c^*=1.6$ and $c/c^*=8.5$. As liquid flows from the droplets  towards the bridge due to capillary pressure, the bridge height grows with time. The corresponding bridge profiles after re-scaling with the similarity parameter have been presented in Fig.~\ref{fig:computational}(b). The collapse of data suggests that similar to Newtonian fluid, polymeric droplets also display self similarity at early stages of coalescence. 

We finally attempt to describe the universal bridge profile using thin film lubrication model. In this regard, we appeal to Varma et al. \cite{varma2021coalescence}, who applied the linear PTT constitutive relation to obtain a modified thin film lubrication equation without gravitational body force for polymeric droplets.

\begin{equation}
\frac{\partial h}{\partial t}+\frac{\sigma}{3 \eta_o}\frac{\partial}{\partial x}\left[h^{3} \frac{\partial^{3} h}{\partial x^{3}}+\frac{6 \kappa \lambda^{2}\sigma^2}{5 \eta_o^{2}} h^{5}\left(\frac{\partial^{3} h}{\partial x^{3}}\right)^{3}\right]=0\\
\label{eq:BC}
\end{equation}
Substituting the similarity parameter into Eq.~\eqref{eq:BC} leads to an ordinary differential equation for the similarity profile $ \zeta (\xi) $, given as 

\begin{equation}
\zeta-\xi\zeta'+\frac{(1+\frac{1}{1+Wi})^4}{16V} \left(\zeta^3\zeta'''\right)'+\frac{27\kappa Wi^2}{2560\theta^2V^3}\left(1+\frac{1}{1+Wi}\right)^{10}\left(\zeta^5(\zeta''')^3\right)'=0\\
\label{eq:sim2}
\end{equation}

Here V=0.818809 is a numerical constant for Newtonian fluid\cite{PRL2012} obtained by scaling the coalescence velocity, v, such that 

\begin{equation}
v=V\frac{\gamma}{3\eta}\theta^4
\label{eq:vscale}
\end{equation}

For Newtonian fluid, $Wi=0$ and Eq.~\eqref{eq:sim2} reduces to the one proposed by Hernandez et al. \cite{PRL2012} Next we list down the boundary conditions required to solve Eq.~\eqref{eq:sim2}. As the two droplets are symmetric at $x=0$, 

\begin{equation}
\begin{aligned}
\zeta\left(0\right)=1\\  
\zeta'\left(0\right)=0\\  
\zeta'''\left(0\right)=0\\
\label{eq:bc1}
\end{aligned}
\end{equation}
The far away bridge profile should match the slope of the contact angle $\theta$. Hence, the far away boundary condition is expressed as  
\begin{equation}
\begin{aligned}
\zeta''\left(\infty\right)=0\\  
\label{eq:bc2}
\end{aligned}
\end{equation}

\begin{figure}
\includegraphics[scale=0.39]{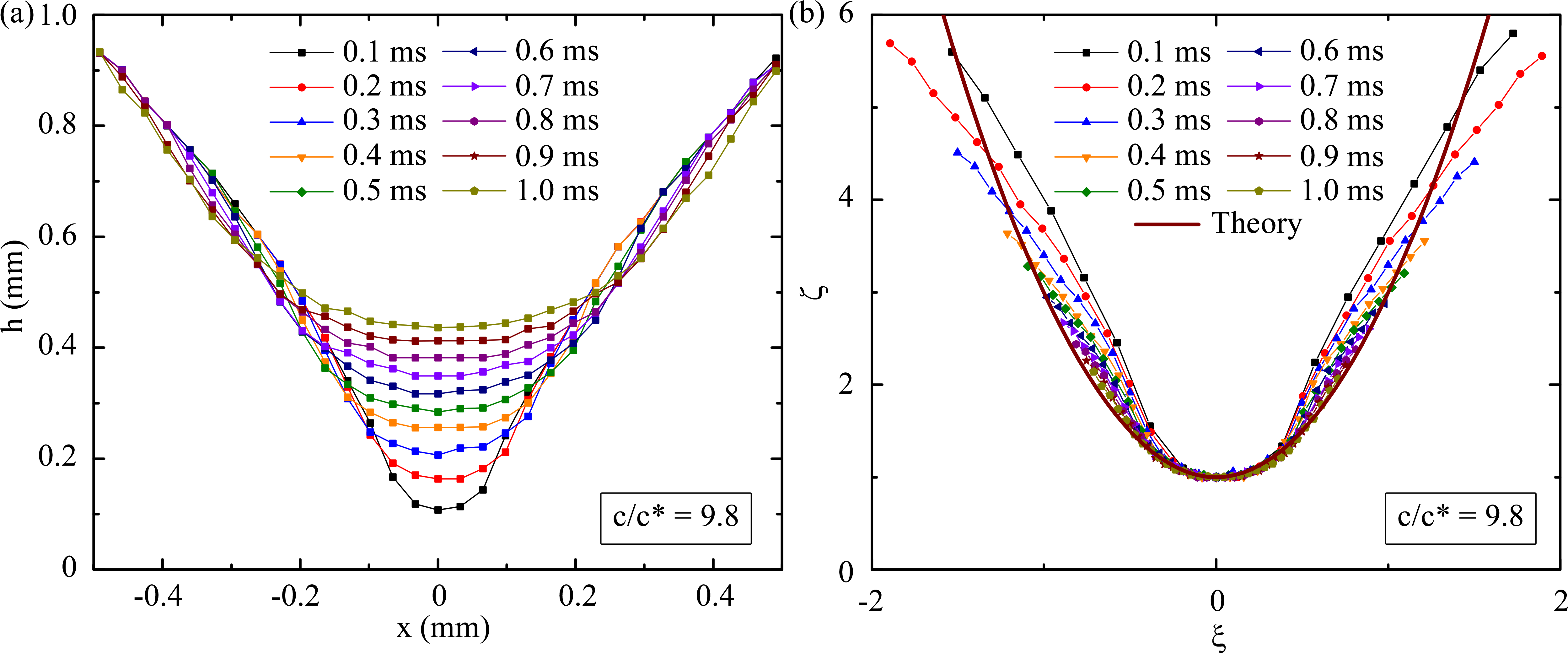} 
\caption{(a) Bridge profiles for concentration ratio $c/c^*=9.8$  at different time instants ranging between 0.1 ms to 1 ms with an interval of 0.1 ms, (b) Comparison between rescaled bridge profile and theoretical similarity solution (solid line) in the same time range}
\label{fig:parametric} 
\end{figure}

Our previous discussion on predominant forces has already established that the thin film approximation is only applicable for the semi dilute entanglement regime $c/c^*>c_e/c^*$. Hence, a concentration ratio of $c/c^*=9.8$, which is higher than the entanglement concentration, has been considered to provide a comparison between the bridge profiles predicted by the lubrication model with that of experiments. Fig.~\ref{fig:parametric}(a) shows the bridge profile for $c/c^*=9.8$ at different instants of time. The rescaled experimental profile along with the theoretical profile (solid line) obtained from lubrication model is presented in Fig.~\ref{fig:parametric}(b). A convincing agreement between the experimental and theoretical profiles suggest that the lubrication model is capable of predicting the coalescence dynamics to an acceptable level. However, some deviation between the experimental and theoretical profiles can be observed which may have resulted due to pinning\cite{chandra2021contact} on the needle surface or the continuous volume growth of the droplets. 
\section{Conclusion}
We unveil that the two methods of merging of droplets, namely, volume filling method (VFM) and droplet spreading method (DSM) leads to dramatically different coalescence dynamics of complex rheology fluids. A discussion on non-dimensional numbers suggest that VFM is inherently associated with strong inertial forces. Moreover, unlike DSM which induces polymer chain elongation, the polymer chains in VFM remain at a relaxed state due to continuous fluid pumping. Consequently, the influence of polymer on the bridge evolution remains quiescent until the polymer chains get highly entangled at large polymer concentrations ($c/c^*>9.8$). The bridge profiles when rescaled with a similarity parameter that incorporates the effect of relaxation time scale, displays a self similar dynamics. Employing the similarity solution in a modified thin film equation for polymers also demonstrates a universal shape of the bridge for $c/c^*>9.8$. However, incorporating the effect of volume growth on polymer chains in the theoretical model can provide better agreement with experimental results and warrants future investigation. The present findings provide a road map for potential innovations in industrial applications tuned for volume growth coalescence of polymeric droplets.\\

\bibliography{bibfile}

\end{document}